# Functionalization of Amorphous Chalcogenide and Titanium Oxide Layers by Gold Nanoparticles

Sandor Kokenyesi[1], Sandor Biri[2], Csaba Hegedus[3],

Stepan Charnovich[1], Attila Csik[2]

[1]Institute of Physics, University of Debrecen, Debrecen, Hungary
[2]Institute for Nuclear Research, Hungarian Academy of Sciences, Debrecen, Hungary
[3]Faculty of Dentistry, University of Debrecen, Hungary

**Abstract.** The technology problems of fabricating different, nanometers sized gold particles in the layered composites like light-sensitive chalcogenide glass/gold nanoparticles/transparent substrate or titanium/titanium oxide/gold nanoparticles were investigated in our work. Combination of ion implantation, plasma deposition with annealing processes results physical routes for creation of gold nanoparticles in the mentioned structures, which possess plasmon effects. These functionalized structures are planned to use for investigations of optical recording processes, biocompatibility of titanium implants.

**Introduction**

Incorporation of nanoparticles into a matrix can essentially change the properties of both the particle and the host, as well as the characteristics of different processes in the composite. Glass/metal nanocomposites are known among a wide number of composites [1]. In particular, gold nanoparticles (GNP) may be incorporated into a matrix and be applied due to the inertness and plasmon resonance effects, which are easily observable in the visible spectral range [2]. Dielectric/metal nanocomposites have been exploited in improved photocatalysts, photochromic and electrochromic films, non-linear optical materials and devices, etc. [3-4]. The influence of plasmon fields on the photoinduced processes in semiconductor chalcogenide glasses was established in [5]. The application of GNPs in medicine also is known, for example in tumor treatments or due to the antibacterial properties [6].

These nanocomposites can be fabricated either by chemical methods, where GNP are formed directly in the composite due to the selected chemical reactions in the mixture of raw components, or by physical methods, like Ostwald ripening [7] of thin gold layers or different sputtering, evaporation, implantation methods combined with further heat treatment. We applied physical routs of GNP fabrication and creation of functionalized nanocomposite structures, based on Ti, usually used for medical implants, and chalcogenide glasses, used for optical recording processes and photonic elements.

**Experimental**

In our work we used 1 mm thick silica glass or Ti substrates on which the GNP and further structures were created. The first step of main interest was the fabrication of GNP with necessary dimensions. In the case of silica glass substrates the process was performed by thermal treatment at 550 C and a normal atmosphere conditions of 15-50 nm thick Au layers, previously deposited by Ar-plasma sputtering, The substrates were chemically and plasma-cleaned before the deposition (the input Ti plates were polished in a normal atmosphere, the glass substrates were normal microscope plates). To get more sophisticated nanocomposites, possibly with enhanced stability and biocompatibility, implantation of Au into the Ti surface before the gold layer deposition was performed with a special ECRIS complex (Electron Cyclotron Resonance Ion Source) at the Institute for Nuclear Research (ATOMKI, Debrecen) [8]. Besides the specific purpose of ECRIS (to provide highly charged ion plasmas and gas ion beams ) it was upgraded for producing Au ion beams with charge states up to +27, what can be used for smoother variation of the surface roughness, depth profiling and stability.

Investigations of composite structures were performed by SEM (Hitachi S-4300), EDAX, AFM (Veeco), the optical characteristics were measured by Shimadzu UV 3600 spectrophotometer.

**Results and discussion**

As far we were interested not only in fabrication of GNP on a given solid substrates, but in obtaining structures where localized plasmon fields generation can be realized and their influence on the selected process can be investigated, the size and shape of the GNPs was essential. The input parameters of the used physical rout were the thickness of the initial continuous gold layer and the temperature and time of annealing. The films were beaded by heat treatment at 550°C for 1-16 hours for both type of substrates. The continuous layer was disintegrated into islands, then during Ostwald ripening the GNPs were formed. The typical optical transmission spectrum of GNP-covered glass substrate, which demonstrate the plasmon resonance absorption and the dependence of absorption maximum on the average size of GNPs are presented in Fig.1.

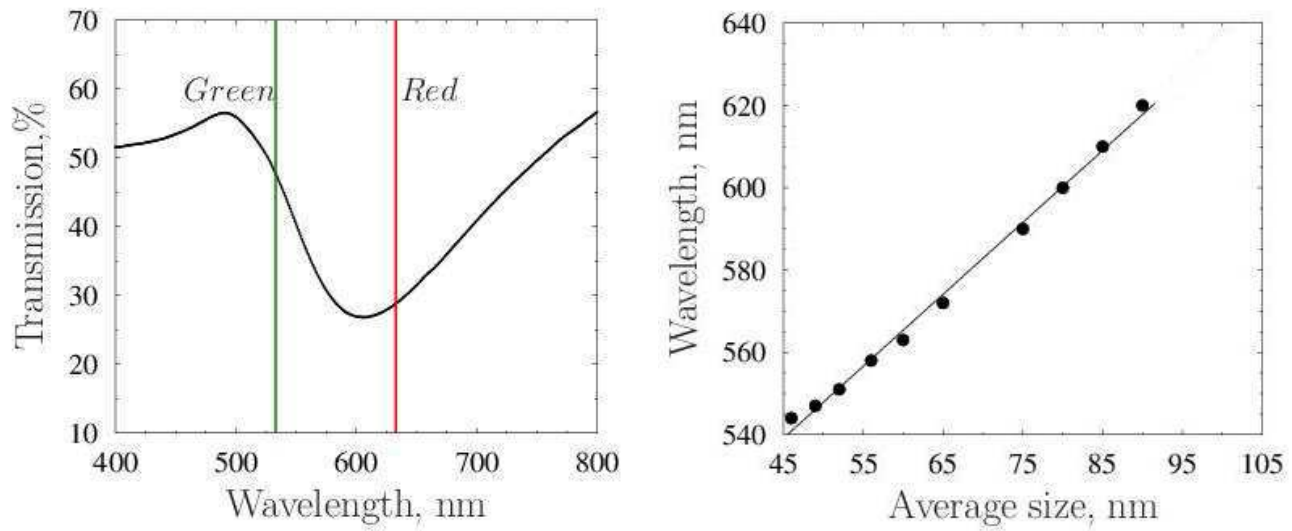

Fig.1. a) Optical transmission spectrum of GNP layer/glass substrate for GNP average size 80 nm (the emission wavelengths of the lasers most frequently used for excitation of nanocomposites are marked). b) Dependence of plasmon resonance maximum on the average size of the GNPs on the glass substrate.

300-500 nm thick amorphous chalcogenide layers were deposited by vacuum evaporation onto the glass substrates covered by GNPs with selected dimensions. It is known that the amorphous As-Se layers undergo light-induced structural transformations, which result in the change of optical parameters and thickness, and these changes can be enhanced in the presence of plasmon fields [5]. So the correlation of the maximum sensitivity ( in the spectral range of optical absorption edge for the given chalcogenide) and the plasmon resonance maximum in such a functionalized structure allows to create layers with enhanced optical (laser) recording parameters. We have done selection of pairs of optimum GNP-chalcogenide glass composition for optical recording by solid state lasers emitting in green and red spectral range. The difference in the transformation rates with and without plasmon enhancement gives us a new possibility to fabricate optical elements (waveguides, lens arrays) as well as scaling down their dimensions.

The thin $TiO_2$ layer on the surface of Ti substrate also allows the use of above mentioned physical rout for GNP fabrication. As far as the mechanical stability of such structure is also important, we tried also to use the thermally induced agglomeration of Au atoms preliminary implanted to the thin, about 20-30 nm thick surface layer on Ti substrate, for GNP creation. The SEM pictures in Fig.2.d) show that the GNPs are created more dense and homogeneously on the surface implanted by Au ions before the plasma sputtering of the Au layer, which undergoes Ostwald ripening during the annealing.

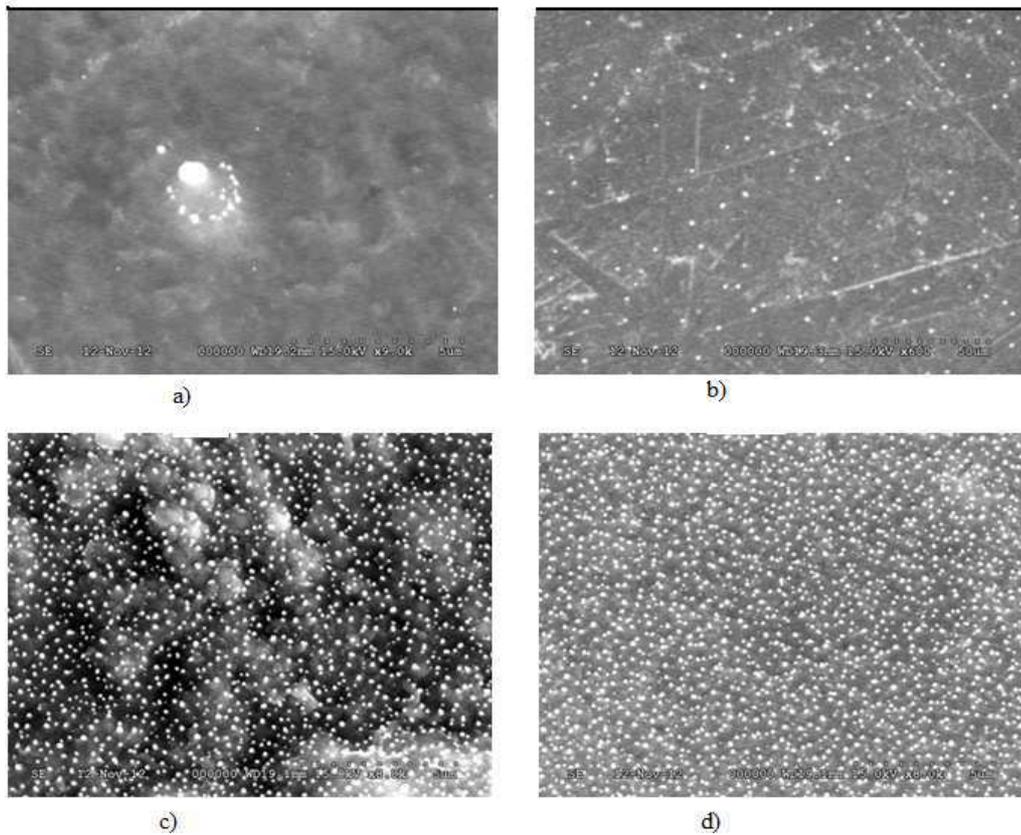

Fig.2. SEM pictures of the annealed Ti surface: without Au (a), implanted with 90 keV gold ions (b), with plasma sputtered 30 nm thick Au layer (c), both implanted and plasma sputtered Au (d).

Probably implanted Au serves additional aggregation centers for surface Au atoms and the total density of GNPs, as well as the smoothness of the surface increase. Plasmon resonance was observed in the reflection spectra of samples d), what enables further experiments on their influence on the composite surface functionalization. Further investigations of depth profiles, bonding strength, as well as of the biocompatibility of these samples with regard to the bone cell growth are in progress.

**Conclusions**

A physical rout of gold nanoparticles fabrication on the surface of glass and Ti substrates was investigated, which allows creation of functionalized surfaces and surface nanocomposites, where plasmon resonances appear under proper illumination. These structures can be used for laser recording of photonic elements, as well as are promising for biomedical applications.

**Acknowledgements**

This work was supported by the TAMOP 4.2.2.A-11/1/KONV-2012-0036 project, which is co-financed by the European Union and European Social Fund.